\documentclass[journal=jacsat,manuscript=article]{achemso}
\newenvironment{authorcontributions}{\section*{Author Contributions}}{}
\newenvironment{funding}{\section*{Funding Sources}}{}

\usepackage[version=3]{mhchem} 

\setcitestyle{numbers,super,sort&compress}

\author{Sinan Genc}
\affiliation[Unknown University]{Department of Electrical and Electronics Engineering, Abdullah G\"{u}l University, 38080 Kayseri, Turkey.}
\alsoaffiliation[Second University]
{Department of Physics, Middle East Technical University, 06800 Ankara, Turkey.}
\author{Oğuzhan Yücel}
\affiliation[Unknown University]{Department of Physics, Middle East Technical University, 06800 Ankara, Turkey.}
\alsoaffiliation[Second University]
{Department of Physics, Freie Universit\"{a}t Berlin, 14195 Berlin, Germany.}
\author{Furkan Ağlarcı}
\affiliation[Unknown University]
{Department of Physics, Izmir Institute of Technology, 35430 Izmir, Turkey.}
\author{Carlos Rodriguez-Fernandez}
\affiliation[Unknown University]
{Department of Physics, Tampere University, 33720 Tampere, Finland.}
\author{Alpay Yilmaz}
\affiliation[Unknown University]
{Department of Physics, Middle East Technical University, 06800 Ankara, Turkey.}
\author{Humeyra Caglayan}
\email{h.caglayan@tue.nl}
\affiliation[Unknown University]
{Department of Physics, Tampere University, 33720 Tampere, Finland.}
\alsoaffiliation[Second University]
{Department of Electrical Engineering, Photonic Integration Group, Eindhoven University of Technology, 5600 MB Eindhoven, The Netherlands}
\author{Serkan Ateş}
\email{serkan.ates@sabanciuniv.edu}
\affiliation[Unknown University]
{Department of Physics, Izmir Institute of Technology, 35430 Izmir, Turkey.}
\alsoaffiliation[Second University]
{Faculty of Engineering and Natural Sciences, Sabancı University, 34956 Tuzla, Istanbul Turkey}
\author{Alpan Bek}
\email{bek@metu.edu.tr}
\affiliation[Unknown University]
{Department of Physics, Middle East Technical University, 06800 Ankara, Turkey.}

\title[An \textsf{achemso} demo]
  {Disorder-Engineered Hybrid Plasmonic Cavities for Emission Control of Defects in hBN}

\begin{document}


\begin{abstract}
  Defect-based quantum emitters in hexagonal boron nitride (hBN) are promising building blocks for scalable quantum photonics due to their stable single-photon emission at room temperature. However, enhancing their emission intensity and controlling the decay dynamics remain significant challenges. This study demonstrates a low-cost, scalable fabrication approach to integrate plasmonic nanocavities with defect-based quantum emitters in hBN nanoflakes. Using the thermal dewetting process, we realize two distinct configurations: stochastic Ag nanoparticles (AgNPs) on hBN flakes and hybrid plasmonic nanocavities formed by AgNPs on top of hBN flakes supported on gold/silicon dioxide (Au/SiO$_2$) substrates. While AgNPs on bare hBN yield up to a two-fold photoluminescence (PL) enhancement with reduced emitter lifetimes, the hybrid nanocavity architecture provides a dramatic, up to 100-fold PL enhancement and improved uniformity across multiple emitters, all without requiring deterministic positioning. Finite-difference time-domain (FDTD) simulations and time-resolved PL measurements confirm size-dependent control over decay dynamics and cavity-emitter interactions. Our versatile solution overcomes key quantum photonic device development challenges, including material integration, emission intensity optimization, and spectral multiplexity.
\end{abstract}


\section{Introduction}
Quantum emitters are central to quantum technologies due to their intrinsically non-classical light emission \cite{fox}. Emitters based on two-dimensional (2D) materials offer a promising platform for scalable photonic quantum technologies, owing to their compactness, stability, and tunable properties \cite{kianinia3}. Among these materials, hexagonal boron nitride (hBN) offers several advantageous properties, such as high chemical and thermal stability \cite{Grosso,herzig}, a wide bandgap (6 eV), and the ability to host room-temperature single-photon emitters based on atomic defects \cite{Yamamura,prasad,Tran2,cakan}. Despite some challenges, such as defect control, spectral variability, and quantum yield limitations, advances in material synthesis, defect engineering, and nanofabrication are progressively overcoming these hurdles, paving the way for the integration of hBN into next-generation quantum photonic devices \cite{hennessey,milad,chili}. Overall, these properties also make hBN an attractive material for cavity systems. Its wide bandgap effectively isolates quantum emitters from environmental noise, enhancing emission stability, while its ultra-thin, atomically smooth layers enable precise control over cavity geometries \cite{KimSejeong,shandilya}. However, similar to the other 2D materials, the main challenges of using hBN as an emitter in a cavity system are the precise defect placement and uniform layer thickness, which remain significant technical hurdles \cite{tran,mendelson}. Additionally, integrating hBN with metallic or dielectric substrates requires advanced fabrication techniques to preserve its intrinsic properties, adding complexity to the manufacturing process. 

The idea of coupling a quantum emitter with a photonic cavity originates from modifying spontaneous emission by changing the environment of the quantum emitter (Purcell effect \cite{Purcell}), which can be used to adjust its emission characteristics. After the first experimental demonstration of modified emission from a single quantum system with a millimeter-sized cavity and with the advances in microfabrication techniques, microcavities came forth due to their comparatively diminutive volumes in the micron range, with the potential to integrate into large-scale photonic circuitry \cite{MEGA}.

\begin{table*}[!t]
\caption{Some examples of cavity-enhanced hBN photoluminescence (PL) studies.}
\label{Table1}
\centering
\begin{tabular}{p{2.5cm} p{11cm} p{1.5cm}}
\hline
Enhancement Factor & Precise Location of hBN to Cavity/Nanostructure & Ref. \\
\hline
11 & Emitters are located at the center of a reflective dielectric cavity for optimized reflection. & \cite{zeng} \\
3 & hBN emitters positioned near gold (Au) nanospheres using atomic force microscopy. & \cite{Nguyen} \\
1.4--4.0 & Emitters transferred from dielectric to metallic substrates to form a hybrid metal--dielectric structure. & \cite{Gerard} \\
$\sim$4 & Emitters localized ($\sim$80 nm) in exfoliated few-layer hBN flakes on lacey carbon grids (relative to NV centers, extremely bright UV SPE at 4.1 eV). & \cite{bourrellier} \\
$<1$ & Emitters located in suspended single-crystal hBN membranes. & \cite{exarhos} \\
$\sim$10 & Carbon incorporation (MOVPE, MBE, HOPG conversion, ion implantation) tunes PL intensity. & \cite{Mendelson2021} \\
250 & Emitters are placed in a nano-patch antenna cavity tightly coupled to the cavity region. & \cite{Xu} \\
8 & Carbon-enriched hBN layers (hBN:C) transferred onto a SiO$_2$/Si substrate. & \cite{karanikolas} \\
2 & Solution-synthesized Ag nanocubes drop-casted on ion-implanted ultrathin hBN flakes on Si substrate. & \cite{dowran} \\
\hline
\end{tabular}
\end{table*}

Metallic nanostructures featuring nanoscale gaps support hybrid plasmonic modes with ultra-small mode volumes and highly localized electromagnetic fields, making them ideal for exploring light-matter interactions at the nanoscale \cite{Barreda,zhang}.  Despite their potential, conventional cavity designs, such as gaps of plasmonic bowtie antennas and dielectric resonators, encounter significant constraints in practical applications. While bowtie antennas are efficient in field concentration, their fabrication is challenging due to the minuscule gap, and they often lack scalability, making them unsuitable for widespread use \cite{ShurenHu,khalil}. Similarly, dielectric resonators exhibit higher mode volumes and provide less field localization and limited Purcell factor enhancement \cite{dimitriev}. 

Additionally, both plasmonic and dielectric systems often struggle with integration into 2D material systems due to mismatched geometries, reduced stability, and inconsistent electromagnetic coupling efficiencies \cite{YuanMeng_arxiv,zhu}. Alternative antenna designs, including dipole, monopole, and Yagi-Uda antennas, address some limitations but introduce their challenges. Dipole and monopole antennas, while simple and efficient, cannot localize fields at nanoscale volumes compared to plasmonic designs, limiting their effectiveness in enhancing light-matter interactions \cite{Patel}. Although Yagi-Uda antennas offer excellent directional emission and gain, they are challenging to miniaturize for nanoscale systems, complicating their integration with 2D material-based quantum technologies \cite{Maksymov,coenen}. Furthermore, their alignment and coupling requirements demand precise fabrication, increasing overall complexity and cost. 

Table \ref{Table1} lists some of the plasmonic enhancement factors of hBN PL as reported in the literature. In studies concerning cavity-enhanced emission from hBN quantum emitters, enhancement factors (EF) exhibit considerable variation based on the type of cavity and the location of the emitter. Dielectric cavities often produce moderate enhancement factor values (e.g., 11 in \cite{zeng}), but plasmonic structures like nano-patch antennas can attain significantly greater enhancements (up to 250 in \cite{Xu}) owing to intense field confinement and near-field coupling. Conversely, systems dependent on random nanoparticle distribution or weak hybrid interfaces frequently yield modest enhancement factors (EF $<4$), highlighting the significance of accurate emitter-cavity alignment and resonant mode optimization. Similarly, dielectric optical cavities have also been used to enhance the PL of hBN \cite{fröch,haussler}. The recorded enhancement factors by dielectric cavities are comparable to those of the plasmonic nanocavities. One promising technique for integrating emitters with plasmonic nanostructures is the thermal dewetting of noble metal films. In this process, a thin metal film breaks up into nano-islands upon annealing, forming stochastic nanoscale antennas near the emitter locations without requiring lithography or precise placement. These disorder-engineered, self-assembled nano-antennas exhibit coupling to nearby defect emission in hBN flakes with a minimal fabrication effort.

In this study, multilayer hBN flakes dispersed in solution were used, drop-casted onto the target substrates. This choice was motivated by the ease of integration, scalability, and compatibility with various substrate types, in contrast to exfoliated or CVD-grown flakes, which often require more elaborate transfer or growth protocols. Drop-casting enables rapid prototyping of emitter–cavity configurations without complex positioning, aligning with our goal of exploring lithography-free plasmonic architectures for quantum photonic applications. While this approach sacrifices some control over flake thickness and orientation, it provides a pragmatic pathway towards device concepts that can be implemented in less resource-intensive fabrication environments. Therefore, we explore two complementary approaches to control the emission from single defects in hBN using plasmonic cavities: (i) direct formation of AgNPs via thermal dewetting of a Ag film on hBN flakes dispersed on a Si substrate and (ii) fabrication of hybrid plasmonic nanocavities by thermal dewetting of a Ag film on hBN flakes dispersed on $\rm Au/SiO_{2}$ coated Si substrates. We investigate the impact of these plasmonic configurations on key emission properties such as photoluminescence intensity, lifetime, saturation behavior, and single-photon purity through a combination of experimental measurements and FDTD simulations. We remark that identifying the exact microscopic defect configuration responsible for each emission line in hBN is beyond the scope of this study. Given the well-known heterogeneity of hBN emitters, our focus is on experimentally demonstrating scalable plasmonic coupling and decay-dynamics control, supported by numerical simulations, rather than assigning a unique atomic defect species to each emitter.

\section{Results and discussion}

Thermal dewetting inherently produces a distribution of nanoparticle sizes and spacings, leading to variations in emitter–nanoparticle coupling strengths. While such stochasticity can cause local fluctuations, reproducible average trends have been established under fixed dewetting conditions \cite{thompson,RUFFINO}, with the $\sim$100$\times$ enhancement observed here representing a typical value across multiple sample regions. Likewise, drop-casting yields hBN flakes with varying thicknesses, which can influence emitter–cavity coupling, particularly in plasmonic architectures with small mode volumes. To capture experimentally relevant geometries, representative thin-flake thicknesses in the range of 10--50~nm, consistent with literature \cite{shorny}, were therefore adopted in the simulations. Although the precise thickness of individual flakes was not measured, the consistent observation of fluorescence quenching and enhancement across multiple emitters and sample regions indicates that the reported coupling effects are not dominated by a single, narrowly defined flake geometry.

It is important to note that the color centers studied here correspond to naturally occurring intrinsic point defects in hBN, whose fundamental optical characteristics are not dictated by the flake preparation method. Previous reports have demonstrated comparable single-photon emission properties in exfoliated, CVD-grown, and solution-processed hBN flakes, including similar zero-phonon line features, polarization behavior, and photon antibunching \cite{kianinia3,Tran2,mendelson}. Accordingly, the focus of the present study is demonstrating scalable plasmonic coupling and decay-dynamics control, supported by numerical simulations, rather than deterministic control of flake thickness or defect identity.

\begin{figure*}[!htp]
    \centering
    \includegraphics[width=1\linewidth]{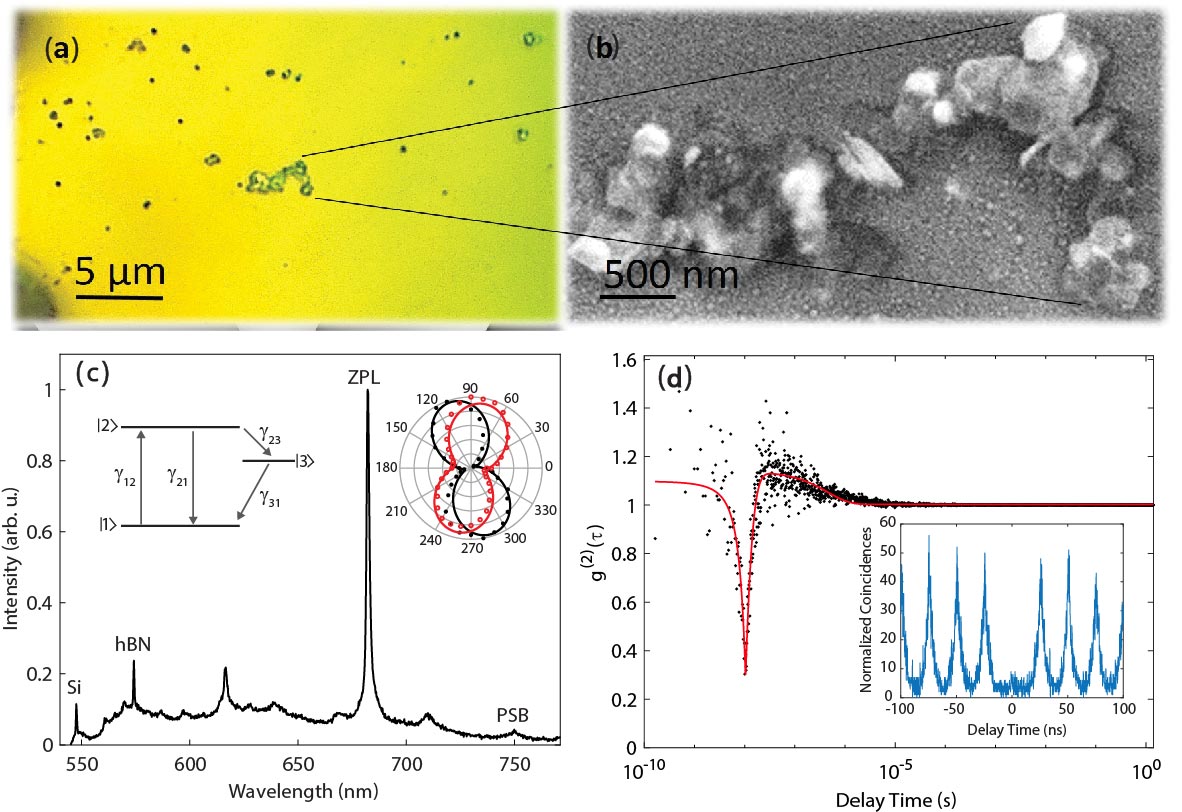}
  \caption{(a) Optical microscope image of the sample surface showing hBN flakes on a silicon substrate. (b) SEM image of selected bulk hBN flakes from the region shown in (a). (c) A representative photoluminescence (PL) spectrum acquired from an hBN flake, showing Raman scattering peaks from both the silicon substrate and the hBN structure, along with a bright zero-phonon line (ZPL) emission from a single defect at 684 nm. (Inset) Polarization dependence of the ZPL emission: excitation (filled circles) and emission (open circles). (d) Second-order photon correlation measurements of the ZPL emission under both continuous-wave (CW) and pulsed excitation at room temperature.}
  \label{Fig1}
\end{figure*}

Figure \ref{Fig1}a shows a typical optical image of the sample surface, where several hBN structures are clearly visible. Figure \ref{Fig1}b shows a scanning electron microscope (SEM) image of the selected bulk hBN flakes, revealing that the drop-casting process results in heterogeneous morphologies, including stacked multilayer regions and significant variations in lateral size and thickness. While such morphology differs from the atomically flat and thickness-controlled flakes obtainable via exfoliation or CVD growth, it reflects a realistic and scalable material platform. While exfoliated or CVD-grown hBN flakes indeed offer superior thickness control and atomically flat interfaces, the present work deliberately employs solution-processed flakes to evaluate plasmonic coupling under fabrication-tolerant and scalable conditions, where perfect 2D interfaces cannot be assumed. In the present work, this morphological variability is not treated as a limitation but serves to examine plasmonic coupling in non-ideal and heterogeneous material environments. To investigate the spectral characteristics of these structures, a home-built micro-photoluminescence ($\mu$-PL) setup was employed. The details of the setup can be found in the Experimental Section.

Figure~\ref{Fig1}c shows a representative room-temperature PL spectrum of a bulk hBN flake excited with a 532~nm continuous-wave (CW) laser. The spectrum exhibits several characteristic features: a Raman peak at 547~nm corresponding to the Si substrate (520~${\rm cm^{-1}}$), and a sharp Raman peak at 574~nm (1370~${\rm cm^{-1}}$) arising from the high-energy $E_{2g}$ phonon mode of hBN. Although hBN also possesses a low-energy Raman-active mode around 52~${\rm cm^{-1}}$, this feature is not visible in our measurements due to strong spectral filtering near the excitation wavelength~\cite{kuzuba,nemanich,Ari_2025}. A sharp emission peak centered at 684~nm, accompanied by a weaker and broader feature at 753~nm, is also observed. These features are assigned to the zero‑phonon line (ZPL)—a sharp emission arising from a phonon‑less electronic transition—and its corresponding phonon sideband (PSB), which originates from emission involving phonon-assisted relaxation. This assignment is consistent with prior observations in hexagonal boron nitride, where the ZPL appears as a narrow peak while the PSB manifests as a broader spectral feature \cite{Ari_2025,Denning}. The energy difference between the ZPL and PSB matches well with the energy of the high-energy phonon mode, confirming their assignment. Polarization-resolved measurements (inset of Fig.~\ref{Fig1}c) reveal strong linear polarization in both excitation (filled circles) and emission (open circles), consistent with a single dipole transition~\cite{2015_Tran,jungwirth,Kumar2024}.

To confirm the quantum nature of the emission, the second-order photon correlation function $g^{(2)}(\tau)$ was measured using a Hanbury Brown and Twiss (HBT) interferometer consisting of a 50:50 beamsplitter and two single-photon counting modules (SPCMs), with
timing recorded via a multichannel time- stamping unit. Here, a ``single defect'' refers to an optically isolated individual emitter within the confocal excitation and collection volume, rather than a deterministically created or structurally isolated atomic defect. Figure~\ref{Fig1}d shows the photon correlation histogram recorded under continuous-wave (CW) excitation at room temperature. The pronounced antibunching dip at zero time delay, with $g^{(2)}(0)=0.3$, confirms sub-Poissonian photon statistics and thus single-photon emission from an individual emitter.

The bunching behavior observed at longer delay times is characteristic of a three-level system, which is commonly reported for solid-state quantum emitters, including defects in hBN~\cite{2015_Tran} and diamond~\cite{kurtsiefer}. A schematic representation of the corresponding energy-level structure is shown in the inset of Fig.~\ref{Fig1}c, comprising a ground state $\lvert 1 \rangle$, an excited state $\lvert 2 \rangle$, and a metastable shelving state $\lvert 3 \rangle$. The relevant transition rates are denoted as follows: \( \gamma_{12} \) (excitation), \( \gamma_{21} \) (radiative \( \gamma_{r} \) and nonradiative \( \gamma_{nr} \) decay), \( \gamma_{23} \) (shelving), and \( \gamma_{31} \) (deshelving). These rates govern the dynamics and intensity of the ZPL emission. The observed deviation of $g^{(2) }(0)$ from the ideal value of zero is attributed to imperfect spectral filtering of the zero-phonon line and the finite temporal resolution of the HBT setup (approximately 400~ps), which limits the ability to fully resolve fast decay processes~\cite{2015_Tran}. To further validate the single-photon character of the emission, photon correlation measurements were also performed under pulsed excitation at 483~nm. Under these conditions, a stronger suppression of the zero-delay peak is observed, yielding an antibunching value of $g^{(2)}(0)=0.1$, as shown in the inset of Fig.~\ref{Fig1}d.

\begin{figure*}[!tp]
    \centering
    \includegraphics[width=1\linewidth]{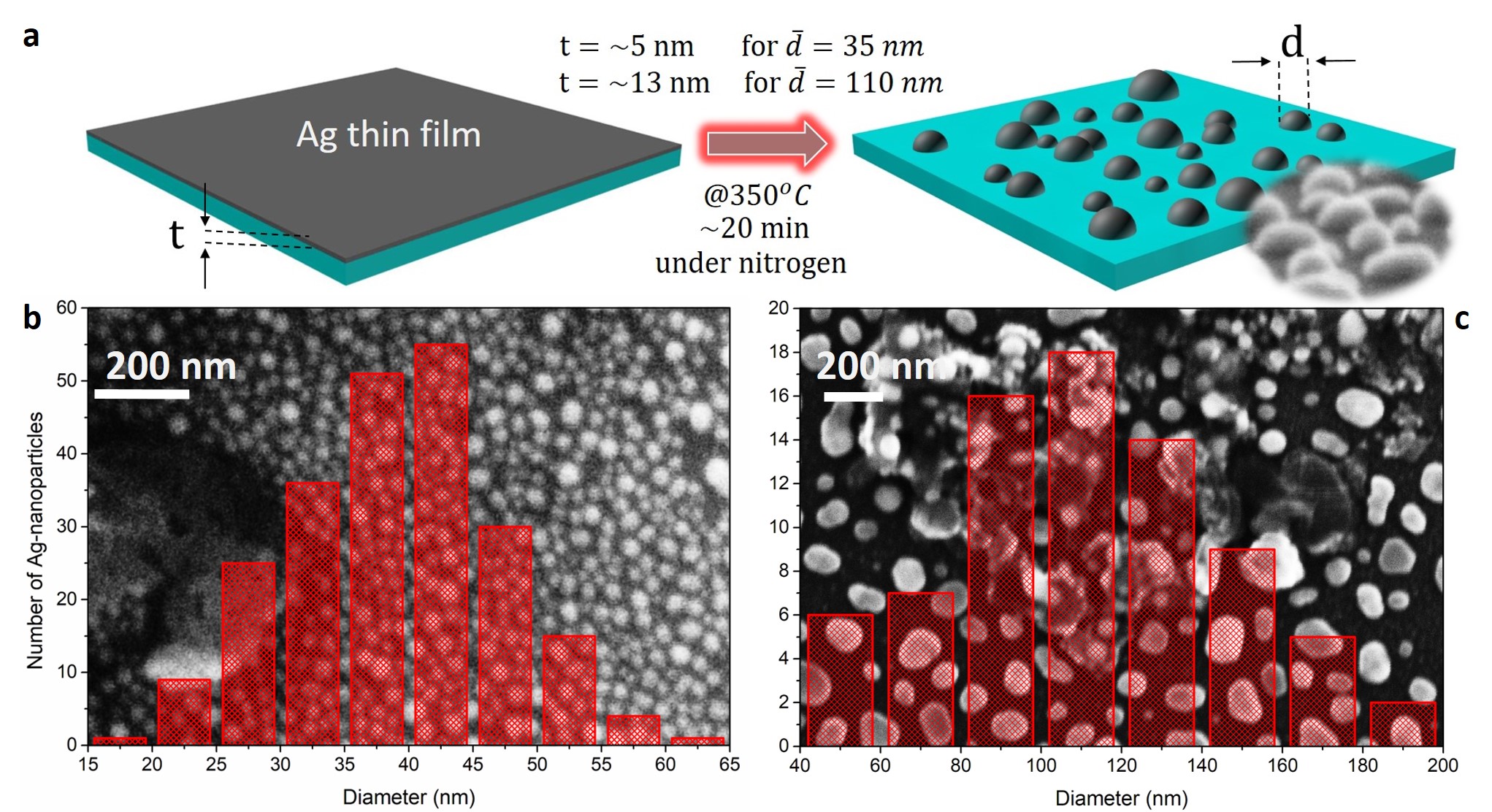}
\caption{(a) Schematic illustrating the fabrication of hemispherical Ag nanoparticles via dewetting of a Ag thin film on a $\rm SiO_2/Si$ substrate. (b) Size distribution histogram and corresponding electron microscope image of smaller AgNPs with an average diameter around 40 nm, used for fluorescence quenching. (c) Size distribution histogram and corresponding electron microscope image of larger AgNPs with an average diameter around 110 nm, used for fluorescence enhancement.}
\label{Fig2}
\end{figure*}

\subsection*{AgNPs on hBN: Emitter Characterization and Plasmonic Coupling}

The nanoparticle sizes investigated here ($\sim$ 35 nm and $\sim$ 110 nm) were chosen to explore two distinct plasmonic regimes. Silver was selected as the plasmonic material due to its superior optical performance in the visible spectral range, where hBN defect emission occurs. Among noble metals, Ag exhibits lower intrinsic optical losses and supports sharper localized surface plasmon resonances compared to Au, enabling stronger near-field enhancement and more efficient modification of radiative decay rates. In addition, Ag films exhibit favorable solid-state dewetting behavior at relatively low annealing temperatures, allowing lithography-free and scalable formation of plasmonic nanoantennas with controlled size distributions.
Particles smaller than $\sim$ 30 nm, while supporting strong local fields, typically have lower scattering cross-sections and higher ohmic losses \cite{Jain2007}, which can limit the radiative contribution to PL enhancement. On the other hand, particles significantly larger than $\sim$ 100 nm can exhibit multipolar plasmon modes and increased radiative damping \cite{Kelly}, reducing spectral overlap with the emitter’s ZPL and altering cavity mode profiles. Optimization across a broader size range may further tailor enhancement for specific applications.

\begin{figure*}[!tp]
    \centering
    \includegraphics[width=0.85\linewidth]{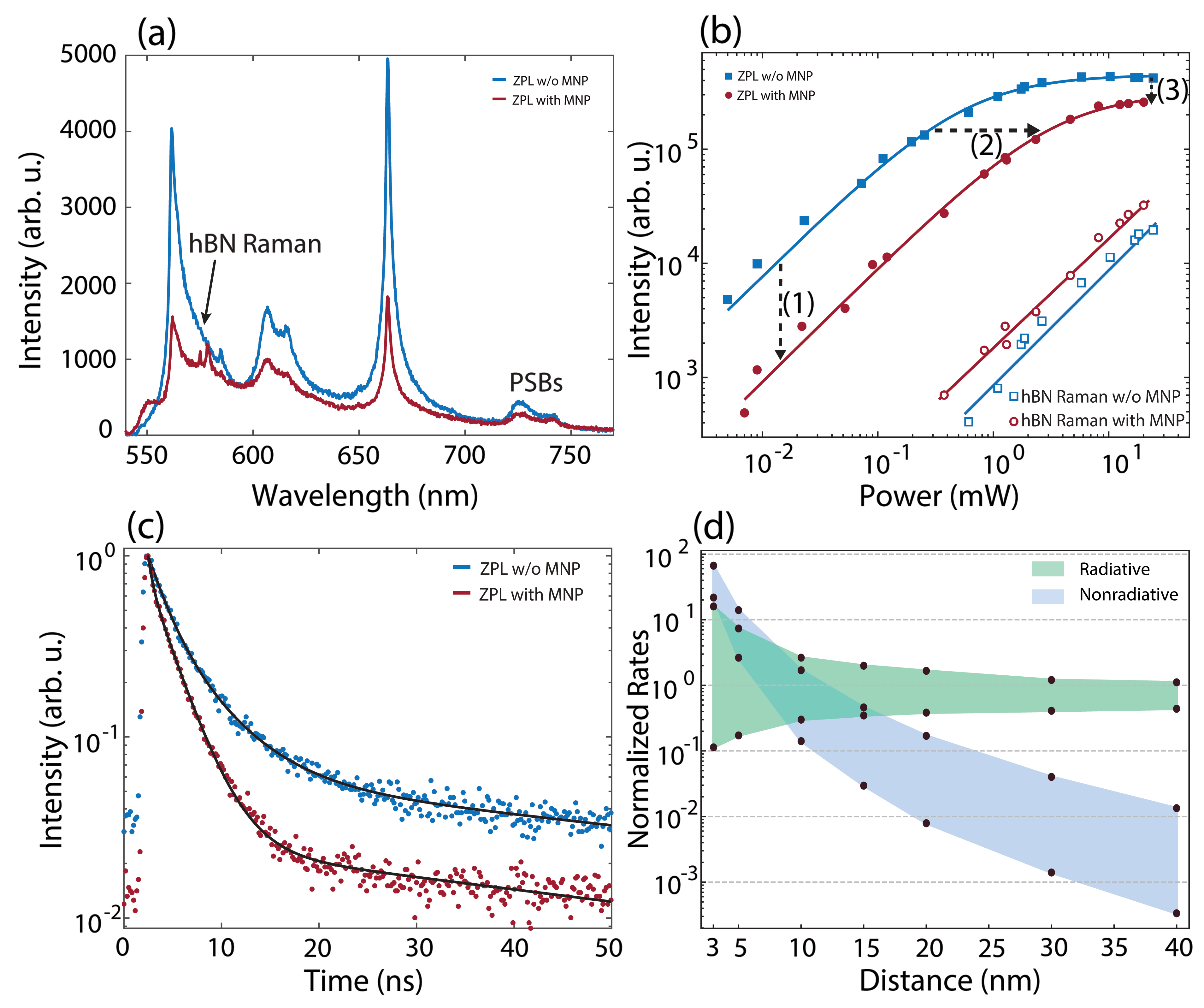}
\caption{(a) PL spectra of a single defect in hBN with a ZPL at 663 nm, measured before and after the deposition of Ag nanoantennas. PSBs appear around 730 nm, while the peak at 574 nm corresponds to Raman scattering of hBN. A significant quenching of ZPL intensity is observed after Ag deposition. (b) Excitation power-dependent intensities of the ZPL and hBN Raman scattering before and after Ag nanoantenna formation. ZPL emission shows strong quenching (are shown with solid circles and squares), while the Raman peak is enhanced due to near-field enhancement effects (are indicated by open circles and squares). (c) Time-resolved PL measurements showing a clear reduction in lifetime after Ag nanoantenna formation, indicating enhanced total decay rate due to coupling with plasmonic modes. (d) FDTD simulation results for a model system consisting of a defect in hBN coupled to a 40 nm diameter AgNP. Radiative and non-radiative decay rates are plotted as functions of emitter–nanoparticle distance. Radiative enhancement dominates at intermediate distances, while non-radiative decay sharply increases at shorter separations. Higher modification is observed for dipoles oriented parallel to the Ag surface, while lower rates occur for orthogonal dipole orientations as shown in the shaded regions.}
\label{Fig3}
\end{figure*}

Figure~\ref{Fig3}a shows PL spectra from a local spot of an hBN sample taken under similar excitation conditions (power and polarization) before and after the fabrication of small AgNPs (35~nm in diameter). The bright peaks at 562~nm and 663~nm are identified as ZPL emission from two defects, and the broad doublets around 610~nm and 740~nm correspond to their associated PSBs, arising from coupling to optical phonons of the host hBN—specifically, the Raman-active phonon at 1370~${\rm cm^{-1}}$ and the IR-active phonon at 1600~${\rm cm^{-1}}$. A small peak at 574~nm appears only in the spectrum with AgNPs and is attributed to the Raman scattering of hBN. The influence of AgNPs is clearly observed as strong quenching in the emission from both defects, affecting both ZPLs and their PSBs. This quenching is particularly pronounced for the ZPL at 562~nm, likely due to better spectral overlap with the localized surface plasmon resonance of the AgNPs. The observed reduction in emission intensity is attributed to enhanced non-radiative decay channels introduced by the proximity of the AgNPs, which facilitate energy transfer from the emitter to the metal nanoparticle. For further analysis, we focus on the defect emitting at 663~nm, as its emission line is more spectrally isolated and thus more suitable for quantitative characterization.

To investigate the origin of the observed quenching in PL emission under identical excitation conditions, power-dependent PL measurements were conducted on the same defect emitting at 663\,nm, both before and after AgNP fabrication. From each spectrum, the intensity of the ZPL emission and the hBN Raman scattering were extracted and are shown in Figure~\ref{Fig3}b for quantitative comparison. While the Raman signal is visible only at high excitation powers, it exhibits a linear power dependence, consistent with its scattering origin. Notably, a clear enhancement in the Raman scattering is observed after the AgNP formation, attributed to surface-enhanced Raman scattering (SERS) \cite{FLEISCHMANN,JEANMAIRE,kneipp}. In contrast, the ZPL intensity shows typical saturation behavior. At low excitation powers, the ZPL intensity $I$ increases linearly with power, following $ I \approx \gamma_{12} \times QE $, where $\gamma_{12}$ is the excitation rate and $QE$ is the quantum efficiency \cite{kurtsiefer,Aharonovich2016}. At higher excitation powers, saturation sets in, and the intensity reaches a maximum limited by the radiative and non-radiative decay rates, as well as the involvement of metastable triplet states. In this regime, the PL intensity can be described by $ I=\gamma_{r}\gamma_{31}/(\gamma_{31}+\gamma_{23})$, where $\gamma_{r}$ is the radiative decay rate, and $\gamma_{23}, \gamma_{31}$ are the intersystem crossing rates associated with the triplet states. The quantum efficiency, defined as $ QE=\gamma_{r}/(\gamma_{nr}+\gamma_{r}+\gamma_{23})$ reflects the competition between radiative decay ($\gamma_r$), non-radiative decay ($\gamma_{nr}$), and intersystem crossing. The significant quenching observed after AgNP fabrication is consistent with an increase in non-radiative decay pathways, likely caused by energy transfer from the emitter to the nearby metallic nanoparticle, which reduces the $QE$.

The saturation profile of the ZPL emission reveals three significant changes upon coupling to the small AgNPs. First, under weak excitation conditions, the ZPL intensity exhibits a quenching factor of approximately 8.4, indicating a strong reduction in the quantum efficiency of the emitter. This low-power behavior arises from a combination of field enhancement and modifications to the emitter's decay dynamics. Specifically, the enhancement of the non-radiative decay rate outweighs enhancements in both the excitation and radiative decay rates, a well-established effect for small AgNPs near quantum emitters~\cite{LAKOWICZ}. Second, the excitation power required to reach the saturation regime increases by nearly a factor of 6 after AgNP coupling. This is consistent with an increase in the total decay rate, as saturation intensity is reached when the excitation rate $\gamma_{12}$ becomes comparable to the sum of the decay channels. The AgNP-induced modification accelerates the decay processes, necessitating higher excitation power to achieve saturation. Finally, at excitation powers well above saturation, the ZPL intensity still shows a $\sim 40\%$ reduction compared to the uncoupled case. Since the saturation intensity is primarily determined by the radiative decay rate, this observation implies that coupling to the AgNP leads to a reduced radiative decay rate. Together, these results demonstrate that the presence of a small AgNP strongly modifies both radiative and non-radiative processes, ultimately diminishing the quantum efficiency and emission intensity of the defect.

Time-resolved photoluminescence measurements (Figure~\ref{Fig3}c) and FDTD simulations (Figure~\ref{Fig3}d) further corroborate the impact of AgNPs on the emitter's decay dynamics. The decay rate modifications are quantified using time-correlated single-photon counting (TCSPC), with representative decay curves shown in Figure~\ref{Fig3}c. After AgNP formation, the fluorescence decay accelerates noticeably, consistent with an increase in total decay rates due to plasmonic coupling. The decay curves are fitted with a bi-exponential function, yielding two characteristic time constants, $\tau_1$ and $\tau_2$. While these components are often associated with distinct physical processes—such as radiative decay from the excited singlet state and population dynamics involving a metastable triplet state—their exact interpretation requires caution. In a three-level emitter system, both $\tau_1$ and $\tau_2$ are effective time constants that reflect contributions from multiple decay channels, including radiative, non-radiative, and intersystem crossing processes. In this context, the faster component $\tau_1$ decreases significantly from 1.87~ns to 0.80~ns, indicating a substantial increase in the total decay rate of the excited state. The slower component $\tau_2$ also shortens from 5.3~ns to 2.89~ns, suggesting that the dynamics of long-lived states, such as the triplet manifold, are also altered, possibly due to changes in intersystem crossing efficiency or enhanced relaxation through modified photonic or non-radiative channels induced by the nearby AgNPs.

Complementary FDTD simulations further support these findings by calculating radiative and non-radiative decay rates as a function of emitter–nanoparticle separation and particle size. As shown in Figure~\ref{Fig3}d, these simulations illustrate how a nearby AgNP (with a diameter of 40~nm) alters the emitter’s decay dynamics. The shaded regions in the plots represent the range of rate modifications arising from the orientation of the emitter's dipole moment relative to the nanoparticle surface, ranging from a dipole aligned \textit{parallel} to the nanoparticle surface (strongest coupling) to one oriented \textit{orthogonally} (weakest coupling). This variation establishes upper and lower bounds for both radiative and non-radiative decay rate enhancements. At emitter–nanoparticle separations less than 10~nm, strong non-radiative enhancement dominates due to energy transfer into lossy plasmon modes of the metal, leading to fluorescence quenching. These effects are more pronounced for dipoles oriented orthogonally to the nanoparticle surface, which couple less efficiently to radiative modes and more strongly to non-radiative channels. Overall, the simulation results are in good agreement with experimental observations of lifetime shortening and emission quenching. Detailed representative simulation environment is given in Supporting Information Figure S2.

To further elucidate the origin of fluorescence quenching observed for smaller AgNPs, wavelength-dependent absorption and scattering cross sections were calculated using FDTD simulations. As shown in the Supporting Information (Fig.~S4), absorption dominates over scattering for small nanoparticles across the spectral range relevant to the hBN ZPL emission, explaining the enhanced non-radiative decay channels and reduced quantum efficiency.

\begin{figure*}[!tp]
    \centering
    \includegraphics[width=0.85\linewidth]{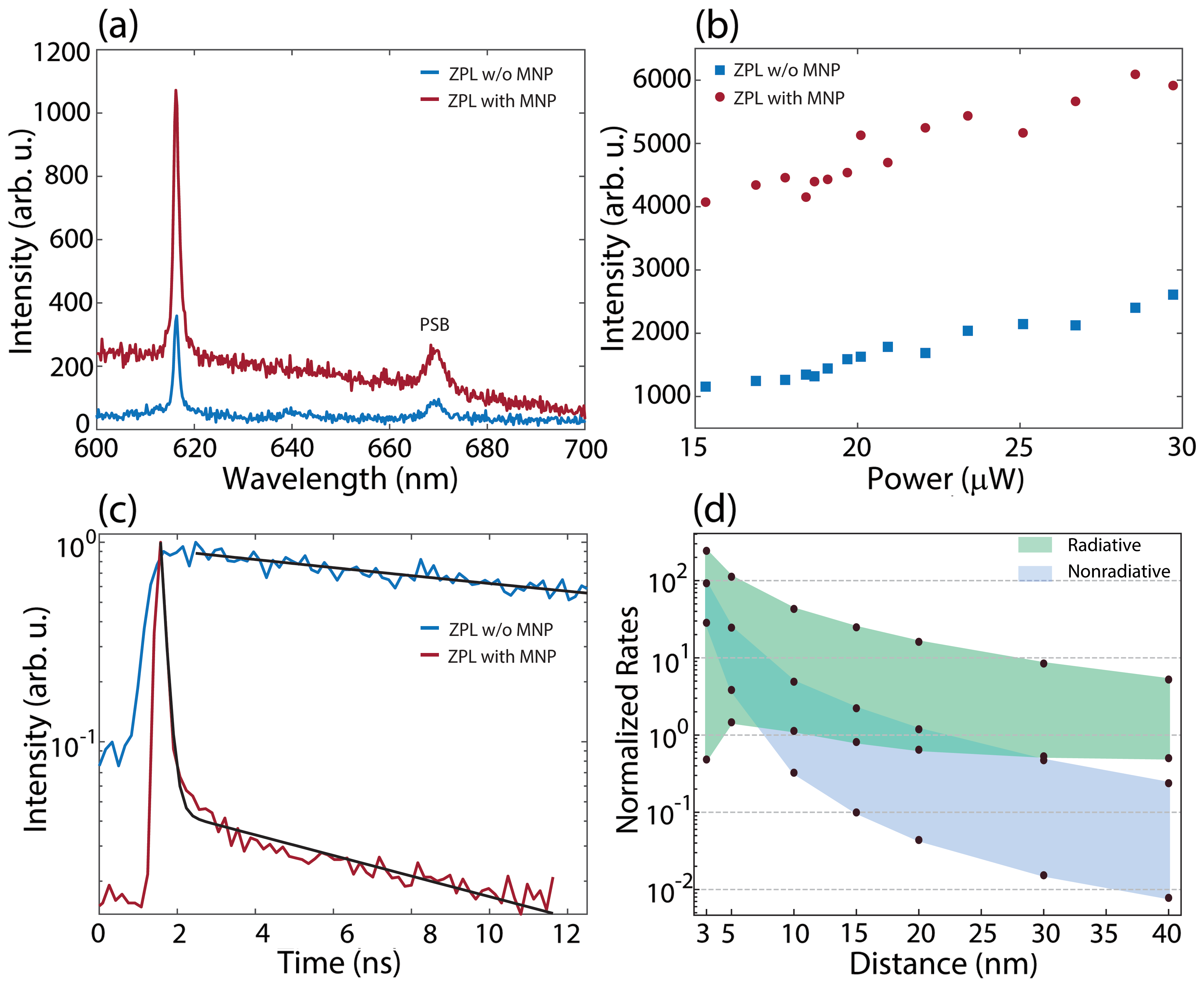}
\caption{(a) PL Spectra of a defect with ZPL at 616 nm before and after the metal nanoparticle taken at similar excitation conditions. Broad peaks around 670 nm are the corresponding PSBs of the defect emission. (b) Excitation power-dependent ZPL intensity before and after the fabrication of AgNPs. A strong fluorescence enhancement is observed. (c) The result of time-resolved measurements on the ZPL emission before and after the AgNPs shows a strong decay time enhancement. (d) Results of FDTD simulations on an ideal system where a defect in hBN is coupled to a  large AgNP with a diameter of 120 nm. Modifications on both radiative and non-radiative rates as a function of emitter-AgNP distance are clearly visible. Simulation parameters were chosen to represent experimentally relevant nanoparticle sizes and emitter–nanoparticle separations; full details are provided in the Supporting Information.}
\label{Fig4}
\end{figure*}

To complement the quenching behavior induced by small AgNPs, we investigate the impact of coupling hBN quantum emitters to larger AgNPs with diameters exceeding 100 nm. Unlike smaller AgNPs, which primarily increase non-radiative losses due to their high absorption cross-sections, larger particles exhibit stronger scattering and support more favorable conditions for radiative enhancement. These nanoparticles can sustain localized surface plasmon resonances that efficiently couple with nearby emitters, modifying the local density of optical states (LDOS) and enhancing the radiative decay channel. As shown in Figure~\ref{Fig4}a, PL spectra obtained before and after AgNP formation reveal a substantial increase in emission intensity from a single defect. The ZPL of the emitter, located at 616~nm, becomes nearly five times more intense following AgNP deposition, and the PSB around 670~nm is significantly more pronounced. These enhancements are indicative of improved radiative recombination pathways facilitated by plasmonic coupling. Importantly, this increase in emission is not accompanied by spectral distortion or shifts in the emission peaks, suggesting that the local electromagnetic environment is modified without compromising the spectral identity of the defect. 

Power-dependent PL measurements, displayed in Figure~\ref{Fig4}b, further support this interpretation. Although the saturation power remains nearly unchanged compared to the pre-coupling condition, the maximum achievable PL intensity increases markedly, indicating an improvement in the radiative quantum efficiency of the emitter. This behavior is in stark contrast to the quenching observed with smaller AgNPs, where both the emission intensity and saturation behavior were adversely affected. In this case, the nearly unchanged excitation threshold alongside enhanced PL intensity implies a selective amplification of the radiative decay rate without a significant increase in non-radiative losses.

Time-resolved measurements shown in Figure \ref{Fig4}c reveal that the ZPL emission at 616~nm initially exhibits a single-exponential decay with a lifetime of $\tau_1 = 20$~ns. After coupling to the large AgNP, the decay becomes bi-exponential, with $\tau_1 = 0.12$~ns and $\tau_2 = 8.4$~ns, indicating a substantial increase in the total decay rate due to strong plasmonic coupling. Despite this dramatic lifetime shortening, the observed PL enhancement is approximately fivefold (Figure~\ref{Fig4}a–b), which appears modest by comparison. This discrepancy arises because lifetime measurements reflect the total decay rate, including both radiative and non-radiative channels, while PL intensity depends primarily on the radiative rate and the emitter's quantum efficiency. Non-radiative losses, changes in emission directivity, and collection efficiency can all limit the net PL enhancement, even when lifetime shortening is substantial. These results demonstrate that large AgNPs can effectively enhance the radiative properties of quantum emitters in hBN, offering a promising route for plasmon-assisted brightness enhancement without significant spectral distortion.

The FDTD simulations, shown in Figure~\ref{Fig4}d, indicate that at emitter–NP separations of approximately 10–25~nm, the radiative decay rate is significantly enhanced, while non-radiative rates remain comparatively low. This favorable regime supports efficient photon emission and aligns well with the experimental PL and lifetime results. Additionally, the enhancement exhibits strong dependence on dipole orientation: emitters with dipole moments parallel to the nanoparticle surface couple more efficiently to the plasmonic mode, resulting in greater Purcell enhancement compared to orthogonal orientations. These results establish a clear nanoparticle size-dependent behavior in the plasmonic coupling of hBN quantum emitters. Engineering the size and geometry of metal nanoparticles thus offers a powerful means to tailor light–matter interactions in two-dimensional quantum photonic systems. Moreover, the observed changes in decay dynamics suggest that plasmonic coupling not only modifies radiative and non-radiative decay pathways, but also perturbs metastable triplet states. This can alter excited-state population dynamics and saturation behavior, consistent with plasmon-mediated control of excited-state kinetics reported in other solid-state emitters~\cite{Wang}.
The FDTD simulations presented here, and given in detail in Supporting Information Figure S3, are intended to capture the dominant physical trends governing plasmon–emitter interactions under experimentally relevant conditions, rather than to provide a one-to-one quantitative match to a specific emitter. Simulation parameters were therefore chosen to reflect the experimentally measured nanoparticle sizes and realistic emitter–nanoparticle separations, while accounting for the unknown dipole orientation of individual defects.

For larger AgNPs, the situation is markedly different. The corresponding wavelength-dependent cross sections (Supporting Information, Fig.~S5) reveal a strong scattering contribution relative to absorption, which favors radiative decay-rate enhancement. This scattering-dominated regime is consistent with the experimentally observed increase in emission intensity and supports nanoparticle size as a key design parameter for engineering radiative versus non-radiative decay pathways.

The contrasting decay-rate modification observed for small and large AgNPs can be understood by considering the spectral overlap between the emitter’s ZPL and the wavelength-dependent absorption and scattering cross sections of the nanoparticles. In the strong coupling regime, where the emitter--nanoparticle separation is small, the dominant decay pathway is governed by whether the ZPL overlaps primarily with absorption-dominated or scattering-dominated plasmonic modes. If the ZPL lies within the absorption band, energy transfer into lossy plasmon modes enhances non-radiative decay and leads to fluorescence quenching. In contrast, spectral overlap with a scattering-dominated response favors radiative decay-rate enhancement, resulting in increased emission intensity. This framework provides a unified interpretation of the quenching observed for smaller AgNPs (Figure~\ref{Fig3}) and the enhancement observed for larger nanoparticles (Figure~\ref{Fig4}).

The FDTD simulations shown in Fig.~4d were performed using geometrical and material parameters selected to reflect experimentally relevant conditions rather than to reproduce a specific emitter–nanoparticle configuration. In the simulations, AgNPs with diameters matching the experimentally observed size distributions were coupled to an electric point dipole embedded in an hBN layer, with emitter–nanoparticle separations swept over the near-field interaction regime. The influence of dipole orientation was accounted for by considering both parallel and orthogonal dipole alignments. Detailed simulation parameters, including geometry, material models, boundary conditions, and numerical settings, are provided in the Supporting Information.

\subsection*{Hybrid Nanocavities on Gold/Silicon Dioxide: Enhanced Uniformity and Confinement}

\begin{figure*}[!tp]
    \centering
    \includegraphics[width=1\linewidth]{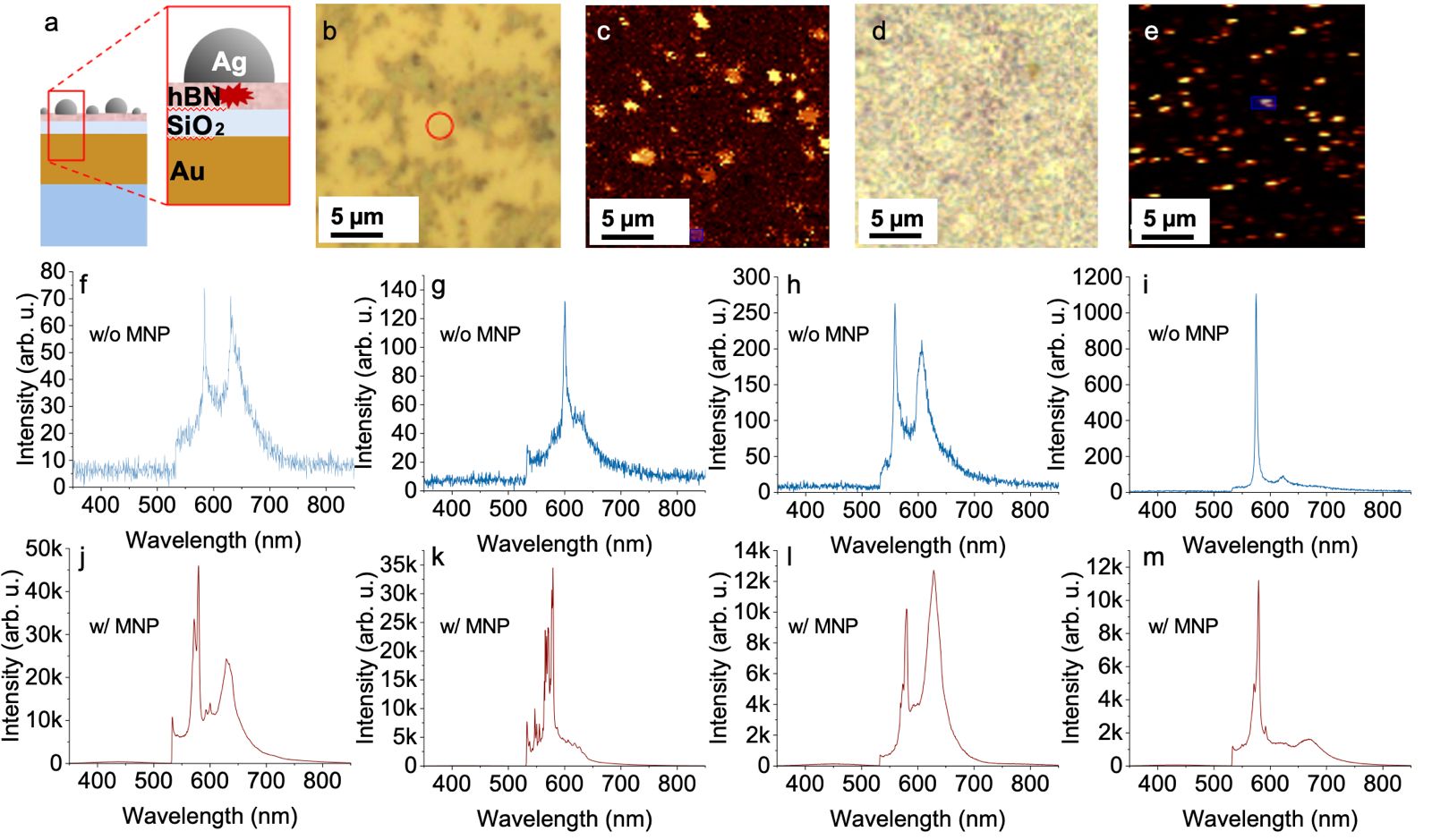}
\caption{ (a) Schematic (not to scale) of the hybrid cavity geometry, indicating silver nanoparticles, hBN flakes, SiO$_2$ spacer layer (20 nm), gold mirror layer (100 nm), and Si substrate. (b, d) Conventional bright-field microscope images with emitter regions marked by red circles (before AgNP formation) and blue circles (after AgNP formation). (c, e) Corresponding confocal laser scanning microscope (CLSM) images; red/blue circles indicate locations of spectra shown in panels (f–m). The apparent distortion in panel (e) is due to slight scanning calibration drift. (f–i) Representative PL spectra from emitters before AgNP formation. (j–m) Representative PL spectra from emitters after AgNP formation.}
\label{Fig5}
\end{figure*}

In the second phase of this study, building upon the emission enhancement observed via thermally dewetted AgNPs, we developed a hybrid cavity platform to further improve the enhancement of emission from defects in hBN nanoflakes. The choice of a SiO$_2$ spacer layer in the hybrid cavity architecture was deliberate and based on optical and practical considerations rather than substrate availability. SiO$_2$ provides optical transparency in the visible range, chemical stability, and a moderate refractive index that enables controlled tuning of the emitter--mirror separation and the local density of optical states. The spacer thickness was optimized through FDTD simulations to balance near-field enhancement and non-radiative losses, making the Au/SiO$_2$ configuration a common and widely adopted platform in plasmonic cavity designs. This platform consists of a metal-dielectric Au/SiO$_2$ bilayer substrate combined with self-assembled AgNPs. Two parameters of the hybrid cavity platform, SiO$_2$ and Au layer thicknesses, are investigated by FDTD simulations, and the thickness values of 20 nm and  100 nm are determined for SiO$_2$ and Au mirror layer for the fabrication process, respectively. Scattering cross-section simulation results for the hybrid cavity with different thickness values can be found in the Supporting Information, Figures S6 and S7. Although thicker SiO$_2$ spacer layers can yield higher absolute scattering cross sections for specific nanoparticle diameters, the experimentally observed emission enhancement occurs predominantly in the 550--650~nm wavelength range. Analysis of the wavelength-dependent absorption and scattering cross sections shows that a SiO$_2$ thickness of 20~nm provides consistent spectral overlap between the hBN zero-phonon line emission and scattering-dominated plasmonic responses across all nanoparticle sizes investigated (70--90~nm). 

It is important to note that tuning the spacer thickness does not represent a trivial optimization of the cavity response to match a specific emission wavelength. The emission spectrum of hBN defect centers is not tunable and exhibits significant emitter-to-emitter variability. Selecting a spacer thickness to maximize scattering at a single wavelength would therefore favor only a limited subset of emitters while compromising broadband coupling and uniformity. The 20~nm spacer thickness was thus chosen to ensure robust enhancement across multiple nanoparticle sizes and emitters, rather than to maximize the response for a single geometry.

The plasmonic nanostructures were formed by the same thermal dewetting method using silver films of varying thicknesses (5--13 nm). The dewetting temperature is significantly lower than the stability threshold of the Au/SiO$_2$ bilayer, the Si substrate, and the hBN flakes, ensuring structural integrity throughout processing. When AgNPs are formed on hBN flakes supported by Au/SiO$_2$ substrates, the resulting vertical nanocavity—comprising the AgNP, hBN layer, and reflective Au film—provides enhanced field confinement and optical feedback. This configuration enables strong plasmonic interactions without complex alignment or lithography, addressing key limitations of conventional cavity designs in terms of scalability and reproducibility.

For comparison, the AgNP-on-hBN/Si configuration discussed earlier (Fig.~\ref{Fig3} and Fig.~\ref{Fig4}) provides one-to-one emitter measurements, illustrating quenching and enhancement through controlled nanoparticle–emitter interactions, whereas the hybrid cavity geometry incorporates an additional Au/SiO$_2$ mirror–spacer architecture that enables stronger field confinement, improved photon extraction, and enhanced uniformity across multiple emitters (Fig.~\ref{Fig5}), which are essential for scalable device concepts. As shown by the representative spectra in Fig.~\ref{Fig5}(f–m), different emitters exhibit ZPLs at distinct wavelengths, highlighting the intrinsically multispectral nature of defect emission in hBN and the broadband response of the hybrid cavity platform.

The reported $\sim$100-fold enhancement should therefore be interpreted as an indicative measure of the broadband enhancement capability of the hybrid cavity platform. It reflects the combined effects of plasmonic near-field coupling, optical feedback from the reflective Au layer, and the stochastic formation of AgNPs, rather than optimized coupling to a single, deterministically positioned defect.

Through plasmonic coupling, the AgNPs significantly enhance emission intensity and modulate the spectral response of hBN emitters. The interaction between the defects, the spacer layers, and localized surface plasmons introduces both field enhancement and spectral reshaping, producing a range of emission profiles depending on the defect type and local environment. These effects manifest as distinct ZPL and PSB features at different wavelengths, reflecting the heterogeneous nature of hBN’s defect states. Importantly, the observed PL enhancement spans the full spectral range of the emitters, underscoring the broadband capability of the proposed platform.

\section*{Conclusion}

This study establishes a powerful and scalable platform for engineering light–matter interactions at the nanoscale by integrating stochastic plasmonic nanocavities with quantum emitters in hBN. Through thermal dewetting of noble metal films, we demonstrate a deterministic-free fabrication route that yields strong and tunable coupling by-design between defect centers and plasmonic nanoantennas achieving photoluminescence enhancements of up to two orders of magnitude. This result firmly establishes these hybrid structures as efficient single photon sources for applications in quantum technologies, i.e. quantum key distribution \cite{wang3,Samaner2022,AlJuboori2023,Tapsin2025}. Our results reveal how local field confinement, decay rate manipulation, and cavity–emitter synergy can be harnessed without complex nanofabrication or precise emitter positioning.

Beyond the quantum photonics implications demonstrated here, the broadband enhancement observed in our hybrid nanocavity platform could be leveraged in other domains. Plasmonic near-field coupling in the presence of nanoscale analytes has been shown to enable label-free single-molecule detection via enhanced light–matter interaction \cite{herkert}. The high field confinement and broadband spectral response also align with requirements for quantum-enhanced sensing schemes \cite{dana}. Furthermore, the solution-processable nature of our approach facilitates integration with biomaterials and microfluidic architectures, creating a pathway towards hybrid photonic–bio interfaces \cite{huimin}. Such properties could also be harnessed for emerging paradigms such as single-photon-controlled neural networks \cite{Mousavi} and optogenetic stimulation \cite{bansal}, where precise, low-energy optical control is essential. While these applications were not experimentally demonstrated in this work, the observed optical enhancements and fabrication scalability suggest their feasibility.

While the present work focuses on a proof-of-principle demonstration, we note that the long-term stability of thermally dewetted AgNPs can be affected by oxidation. Literature reports indicate that dielectric overcoating or encapsulation methods can mitigate these effects and preserve plasmonic performance over extended periods \cite{rycenga,BETANCOURT}. Such strategies may be considered in future studies aimed at practical device implementations.

\section{Methods}

Multilayer hBN flakes were obtained commercially in solution form (Graphene Supermarket) and drop-casted (${\approx 10\ \mu}$L) on a silicon substrate, which has laser engraved marks to locate specific emitters before and after the fabrication of AgNPs. The hBN flakes used in this study were commercially obtained and therefore contain naturally occurring intrinsic defect-based emitters. We did not employ defect creation steps such as ion implantation, irradiation, or high-temperature activation anneals. In this work, the term ``defect'' refers to optically active color centers naturally present in the as-received hBN flakes, which are subsequently coupled to plasmonic nanostructures.

To investigate the effect of Ag nanoantennas on the optical properties of single defects in hBN, the antennas were fabricated in the vicinity of defect sites using a self-organized solid-state dewetting approach, rather than relocating the emitters to pre-patterned structures~\cite{Nguyen}. For this purpose, Ag thin films of 5 or 13 nm were deposited on hBN-coated silicon substrates and subsequently annealed at $\sim350,^\circ$C in a nitrogen environment~\cite{Jiran90,mona,Nasser2013}. This method is gentle on both the hBN flakes and the silicon substrate, which are stable up to $\sim800\,^\circ$C~\cite{Kianinia}. Silver was selected owing to its low optical losses, narrow plasmonic resonances, and favorable dewetting characteristics compared to other noble metals such as gold. The initial film thickness was used to tune the antenna size distribution, yielding average diameters of $\sim$35 nm and $\sim$110 nm for 5 nm and 13 nm films, respectively (Figure~\ref{Fig2}b,c). These regimes were chosen to probe both fluorescence quenching and enhancement in single hBN defects, since Ag nanoantenna dimensions strongly determine their absorption and scattering properties~\cite{LAKOWICZ}.
By controlling the dewetting duration, the size distribution of the resulting nanoantennas can be tuned (Supporting Information Figure S1). During annealing, the Ag film destabilizes well below the bulk melting point due to its high surface-to-volume ratio~\cite{Jiran92}. Diffusion and agglomeration of Ag atoms lead to the spontaneous formation of hemispheroidal nanoislands distributed across both hBN flakes and the silicon substrate (Figure~\ref{Fig2}a). Owing to the stochastic nature of dewetting, a fraction of these nanoantennas form in close proximity to defect centers, enabling emitter–antenna coupling without additional positioning steps.

hBN defect centers are excited by a CW laser ($\lambda_{exc}$=532 nm, Verdi-V6 Coherent) and a pulsed diode laser ($\lambda_{exc}$=483 nm, 65 ps pulse width, Advanced Laser Diode Systems). Fluorescence is collected using a 50X/0.75 NA objective (Optika). The optimum polarization angle of the excitation is determined using a motorized half-waveplate (HWP). In the detection part of the setup, the Rayleigh scattered light is rejected with a 540 nm notch filter. The sample surface is imaged by a CMOS camera white light illumination. A spectrometer (Andor Shamrock 750) is employed with an EMCCD camera (Andor Newton). The lifetime measurement of the quantum emitter is performed by a photon counting avalanche photodiode (APD) mounted on the HBT interferometer. APDs are connected to a time tagging electronic module (TTM8000, Roithner Laser Technik).

For each substrate configuration, background spectra were recorded from adjacent hBN-free regions on the same substrates (including Au/SiO$_2$ $\pm$ AgNPs) under identical acquisition settings and subtracted from emitter spectra prior to calculating ZPL-based enhancement factors; Raman and SERS features were excluded from the ZPL-based analysis. In our study, the “nanoparticle-only” condition is represented by AgNPs on hBN/Si (i.e., without the Au/SiO$_2$ mirror–spacer), which isolates the near-field plasmonic effect and results in either modest enhancement (for large AgNPs) or fluorescence quenching (for small AgNPs), in agreement with expectations. A separate “hBN on Au/SiO$_2$ without AgNPs” dataset is not included, as under our excitation and collection conditions stable single-emitter PL on Au/SiO$_2$ prior to dewetting was not reliably detectable, indicating that the mirror–spacer alone does not account for the strong enhancement reported here. Likewise, “Au/SiO$_2$ + AgNPs without hBN” cannot produce defect-related ZPL or PSB features; instead it serves to gauge background/SERS.


\begin{authorcontributions}

S.G. and O.Y. contributed equally. 
S.G., O.Y., and A.B. conceived the study. 
S.G. and O.Y. performed the experiments and analyzed the data. 
C.R.-F., A.Y., H.C., and S.A. contributed to sample preparation, measurements, and interpretation.
S.G., F.A., and A.Y. contributed to simulations. 
A.B., H.C., and S.A. supervised the project together. 
S.G., A.B., H.C., and S.A. wrote and reviewed the manuscript and all authors approved the final version.

\end{authorcontributions}

\begin{funding}
T\"urkiye Scientific and Technological Research Council (T\"UB\.ITAK), project 118F119. Middle East Technical University Postdoctoral Research Program (DOSAP-B).
Academy of Finland Flagship Programme (PREIN) (320165).
Horizon2020 Marie SkłodowskaCurie grant agreement No. 895369.
\end{funding}

\begin{acknowledgement}

The authors thank Selçuk Yerci for the workstation on which we performed our computations, Vasilis Karanikolas for reviewing some of the calculations, and Melike Şeyma Tuna for help with the figures. S. G. acknowledges the Middle East Technical University Postdoctoral Research Program (DOSAP-B). S. A. acknowledges the support of the Türkiye Scientific and Technological Research Council (TÜBİTAK) under project number 118F119. H.C. acknowledges the financial support of the Academy of Finland Flagship Programme (PREIN) (320165). C.R.-F. acknowledges the Horizon 2020 research and innovation program under the Marie SkłodowskaCurie grant agreement No. 895369.

\end{acknowledgement}

\begin{suppinfo}

Annealing process, FDTD simulation details, Wavelength-dependent absorption and scattering cross sections, Spectral overlap between hBN ZPL and nanoparticle cross sections. 

\end{suppinfo}

\bibliography{reference}

\newpage

\textbf{For Table of Contents Use Only}

Disorder-Engineered Hybrid Plasmonic Cavities for Emission Control of Defects in hBN\\

Sinan Genc, Oğuzhan Yücel, Furkan Ağlarcı, Carlos Rodriguez-Fernandez, Alpay Yilmaz, Humeyra Caglayan, Serkan Ateş, and Alpan Bek\\

A low-cost and scalable strategy for integrating plasmonic nanoparticles onto hBN nanoflakes, enabling strong and reproducible enhancement of defect emission without the need for deterministic positioning or alignment. \\

\begin{figure}[h!]
  \centering\includegraphics[width=8.25cm]{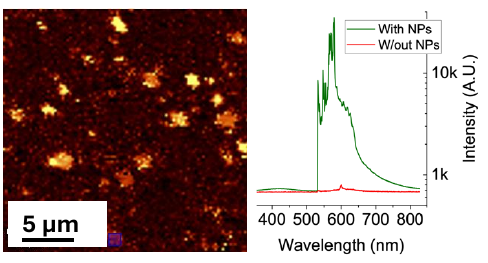}
\end{figure}

\end{document}